\documentclass[numberedappendix,appendixfloats]{emulateapj}

\shorttitle{Velocity Correlation in arbitrary Strength Turbulence}
\shortauthors{Fraschetti, Giacalone}

\usepackage{color}
\usepackage{amssymb,amsmath}
\usepackage{epstopdf}
\usepackage{pstricks,pstricks-add,pst-node}

\newcommand{\vect}[1]{\mathbf{#1}}   

\begin{document}

\title{Early-time velocity autocorrelation for charged particles \\ diffusion and drift in static magnetic turbulence}

\author{F. Fraschetti\altaffilmark{1,2} and J. Giacalone\altaffilmark{1}}
\affil{$^1$Departments of Planetary Sciences and Astronomy, University of Arizona, Tucson, AZ, 85721, USA}
\altaffiltext{2}{Associated Member to LUTh, Observatoire de Paris, CNRS-UMR8102 and Universit\'e Paris VII,
5 Place Jules Janssen, F-92195 Meudon C\'edex, France.}

\begin{abstract}
Using test-particle simulations, we investigate the temporal dependence of the two-point velocity correlation function for charged particles scattering in a time-independent spatially fluctuating magnetic field derived from a three-dimensional isotropic turbulence power spectrum. Such a correlation function allowed us to compute the spatial coefficients of diffusion both parallel and perpendicular to the average magnetic field. Our simulations confirm the dependence of the perpendicular diffusion coefficient on turbulence energy density and particle energy  predicted previously by a model for early-time charged particle transport. Using the computed diffusion coefficients, we exploit the particle velocity autocorrelation to investigate the time-scale over which the particles ``decorrelate'' from the solution to the unperturbed equation of motion. Decorrelation time-scales are evaluated for parallel and perpendicular motions, including the drift of the particles from the local magnetic field line. The regimes of strong and weak magnetic turbulence are compared for various values of the ratio of the particle gyroradius to the correlation length of the magnetic turbulence. Our simulation parameters can be applied to energetic particles in the interplanetary space, cosmic rays at the supernova shocks, and cosmic-rays transport in the intergalactic medium.
\end{abstract}


\section{Introduction}

In several physical systems containing a fluctuating magnetic field in a turbulent plasma, 
the diffusion of charged particles in the direction perpendicular to the average magnetic field
has been recognized to be important. First, in the heliospheric environment, solar energetic particles 
associated with impulsive solar flares, or compact point-like sources, have been observed 
by multiple spacecraft orbiting on the heliospheric ecliptic plane, e.g., STEREO A/B, SOHO, 
widely separated in longitude; this has been interpreted as evidence 
for strong cross-field diffusion \citep{wc06}.
Significant perpendicular diffusion has been also invoked to explain recurrent cosmic-ray variations at very high heliospheric latitudes, possibly connected with Corotating Interaction Regions \citep{kj98}. Observations of solar energetic particles at high heliographic latitude made by Ulysses, have also been interpreted as evidence for cross-field diffusion \citep{z03,d03}. 

Second, in non-relativistic collisionless shocks, e.g., interplanetary shocks or supernova remnant (SNR) shocks, the acceleration rate depends on the shock obliquity \citep{j87}, i.e., the angle between the magnetic field at the shock and the direction normal to the shock surface.
At a perpendicular shock, since the magnetic field lines are  frozen with the plasma flow, the transport perpendicular to the magnetic field lines allows the particles to remain near the shock enhancing their acceleration.
Recent multiwavelength campaigns of SNRs (from radio up to $\gamma$-rays) has not yet constrained the magnetic field obliquity at the SNR shock despite a burgeoning number of observational evidences of the magnetic field amplification.

A single model unifying the transport in the directions parallel and perpendicular to the average magnetic field is needed to understand the propagation of energetic particles in a broad class of environments, i.e. interplanetary space, supernova remnant shocks and interstellar space; however such a model is still missing. 
Perpendicular diffusion has been also studied recently disentangling two different contributions: field line random walk and gradient/curvature drift of the particle guiding center from the local field line \citep{fj11}.
In the limit of weak turbulence and of low ratio of particle gyroradius to the magnetic turbulence coherence length, 
the usual assumption of particle magnetization, i.e., particle following field lines, 
is found to fail even in a simple and idealized turbulence geometry, such as the three-dimensional isotropic model.
For this particular turbulence, this result must {\it a fortiori} hold for high-energy particles, because, due to larger gyroradius, 
high-energy particles decorrelate from the initial field line earlier than low-energy particles.
 
In the present paper we investigate via first-principles, using Monte Carlo numerical simulations, the transport of charged test particles in the direction perpendicular to the average magnetic field at early times, prior to the spatial diffusion phase. The magnetic turbulence induces a decorrelation of the particle velocity from its unperturbed evolution. The two-point particle velocity correlation function, used here indistinguishably from particle velocity autocorrelation, would be constant in time in the absence of field fluctuations. The decorrelation time, defined as the time beyond which the Lorentz force experienced by the charged particle in the ambient magnetic field is uncorrelated with the force initially acting on it, depends physically on the magnetic energy density and on the ratio of the gyroradius to the turbulence correlation length. For a broad range of the parameters studied in this paper, the perpendicular particle velocity decorrelates within a few gyroperiods, i.e., much earlier than the typical diffusion time-scale.

In the numerical experiments performed in the present paper, we relate the time-dependence of the
two-point parallel and perpendicular particle velocity correlation function to the diffusion coefficient as given 
by the TGK formalism \citep{t21,g51,k57}. We explore the regime of strong turbulence and use a broad range for the ratio 
of particle gyroradius to turbulence correlation length; our results apply both to non-relativistic and relativistic particles. Our simulations confirm the exponential, or Markovian, decay of the pitch-angle correlation. However, we find an unexpectedly fast decay of the perpendicular correlation, occurring within a few gyroperiods, which is not accounted for in any phenomenological transport model.

This work is organized as follows: in Section \ref{model} we describe previous models for the velocity autocorrelation; in Section \ref{nummeth} we specify the particular turbulence model and the parameters for the test particles used in our simulations; in Section \ref{diffcoeff} we summarize the results for the behaviour of diffusion coefficient at strong turbulence and high ratio of gyroradius to the turbulence correlation length; we also show that our previous weak-turbulence model based on the separation between field line random walk and gradient/curvature drift is in agreement with our simulations; in Section \ref{velcorr} we discuss our results for the pitch-angle correlation function and the perpendicular velocity autocorrelation; in Section \ref{astro} we describe some astrophysical applications; in Section \ref{summconcl} we summarize and conclude with the observational implications of our results; in Appendixes A and B some details of the numerical code are outlined.

\section{Spatial diffusion models based on velocity autocorrelation}\label{model}

The investigation of the particle motion in times prior to the diffusion phase can be conveniently based on the particle velocity autocorrelation, named Lagrangian as it makes use of the instantaneous particle velocity, well-defined at times as short as the early ballistic phase. Moreover, the velocity autocorrelation offers the most direct approach to the long-range correlation in the so-called anomalous transport regime \citep{ba04}. The Lagrangian velocity autocorrelation has not been frequently applied to particle transport for two reasons: 1) experimentally it is simpler to measure the Eulerian velocity autocorrelation, relating two points in space separated by a fixed coordinate distance, than the Lagrangian velocity autocorrelation, relating the velocity of the same physical particle at two different times; 2) numerical simulations require using a large number of particles to obtain good statistics.
 
The random walk of charged particles in a static, i.e., time-independent, turbulent magnetic field is commonly represented as 
a sequence of many stochastic events, independent of one another, i.e., 
a compound Poissonian process \citep{f71}.
In a magnetic turbulence, every scattering encounter has a probability of {\it not} occurring exponentially decreasing in time, since a scattering event will certainly occur as the time proceeds. An exponential decay 
implies that every scattering event is independent from any other and the transport 
is endowed with complete lack of memory (Markovian process).

An exponential form for the autocorrelation of particle velocity $v(t)$ moving in a turbulent fluid 
($ \langle v(t) v(0)\rangle = v_0^2 e^{-t/T}$) was first heuristically used by \citet{t21} 
in a seminal discussion of the diffusion coefficient in the absence of magnetic field; 
here $T$ has the meaning of characteristic time-scale beyond which the fluid density variations or turbulent fluid motions 
cause a jump of the particle velocity to a value uncorrelated with the previous value.
If magnetic turbulence is included, more recent theoretical arguments led 
to the exponential form of the particle velocity correlation function in the direction 
along the average magnetic field, or, equivalently, pitch-angle correlation function 
$\langle \mu(0) \mu(t) \rangle = 3 \langle v_z(0) v_z(t) \rangle / v_0^2$ 
in a static magnetic field \citep{e74}.
The exponential form of $\langle \mu(t)\mu(0)\rangle$ has been shown \citep{f77} to correspond 
to the closed form for the pitch-angle coefficient diffusion in \cite{j66} for a quasi-isotropic scattering.

We briefly recall some previous approaches to the particle velocity autocorrelation to estimate 
the diffusion coefficients $\kappa$. In weak turbulence regime and
for particle isotropically scattering in all three space directions, 
the pitch-angle correlation is found to decrease exponentially in time \citep{f77}, 
i.e., the scattering is Markovian, with scattering time-scale $\tau_{\parallel}$ 
related to the parallel mean free path ($\lambda_{\parallel}$) by $\lambda_{\parallel} = v \tau_{\parallel} $ and
to the parallel coefficient diffusion by $\kappa_{\parallel} = (v^2 /3) \tau_{\parallel}$. 
In the presence of strong turbulence, numerical simulations \citep{clp02} found
an empirical scaling of the scattering frequency ($1/ \tau_{\parallel}$) with the turbulence magnetic energy density.
This scaling is based on two assumptions: 1) the motion perpendicular to the local field line, 
and the eventual decorrelation from the field line at high rigidity, is dominated by the turbulent scales smaller 
than the gyroradius; 2) the decorrelation time $\tau_\perp$ is smaller than the scattering time $\tau_\parallel$. 
In \citet{fj11}, the scale separation in the former assumption 1) of \citet{clp02} 
is found to hold only for slab turbulence: the decorrelation from the field line is governed by the scales smaller than
the correlation length, but much larger than the gyroscale (it should be noted also that \citet{fj11} assume a negligible power 
in the magnetic turbulence at the gyroscale). In contrast, for a 3D-isotropic magnetic fluctuation, \citet{fj11} found that 
the decorrelation from the local field line has equal contributions from turbulent scales both larger and smaller than the correlation length.
Secondly, in a strong turbulence the fluctuating magnetic field transversal to the average field becomes comparable or larger than $B_0$ and the assumption 2) in \citet{clp02} might be violated.

\cite{bm97} proposed a model (hereafter BAM model) for the perpendicular transport of high-energy (a few $GV$ for typical interplanetary conditions at $1$ AU) charged particles 
based on a specific {\it ansatz} on the form of the perpendicular particle velocity autocorrelation, 
by supposing that particles simply move along the field lines.
Numerical simulations \citep{gj99} have shown that the standard quasi-linear theory \citep{j66} provides a diffusion coefficient in the perpendicular direction larger than its numerical estimate at protons energy between $1$ MeV and $100$ MeV, whereas the diffusion coefficients in classical scattering theory \citep{fg75} or the BAM model are too small.
A more recent theory for perpendicular transport closer to numerical findings than the two forementioned models 
is the NLGC model \citep{mqbz03}. For magnetized plasma flows, NLGC relies on an approximate relation \citep{c59}
between the Eulerian and the Lagrangian velocity autocorrelations. 
The Corrsin relation assumes a diffusive nature of the particle displacement between the two points in space 
where the correlation is computed. However, such a diffusive assumption has been numerically found to hold  
only for purely hydrodynamic flows as in \citet{k77} but not for MHD flows, as far as we are aware.
Moreover, the NLGC, being intrinsically diffusive, can not apply to non-diffusive magnetic turbulences geometries: 
different turbulence geometries are found to have different effects on the particle diffusion (see for example \citet{fj11}). 
In addition, the field line random walk relies on a free parameter (named ``a'' in \citet{mqbz03}) 
which is not provided by any other auxiliary model and must be empirically determined from numerical simulations.
Within the idealized compound diffusion model, where particles are strictly tied to magnetic field lines, 
but scatter along them and trace back along the same field line, the temporal evolution of the perpendicular velocity autocorrelation 
and the consequent diffusion coefficient have been determined analytically, confirming the expected subdiffusion \citep{kj00}.

The velocity autocorrelation $ \langle v_i(t) v_j(0)\rangle$ is known to be related to the symmetric part 
of the instantaneous diffusion tensor $\kappa_{ij} (t)$ by the TGK formalism: 
$\kappa_{ij} (t) =   \int_0^{t} d\xi  \langle v_i(\xi) v_j(0)\rangle $. The heuristic and physically motivated 
BAM forms of the perpendicular velocity autocorrelations at gyroperiod scale are:
$R_\perp (t) = \langle v_x(t) v_x(0) \rangle \propto (v^2 /3) e^{-t/\tau_{\perp}} \cos(\Omega t)$
and $R_A (t) = \langle v_x(t) v_y(0) \rangle \propto (v^2 /3) e^{-t/\tau_{A}} \sin(\Omega t)$,
where $\Omega = e B_0/m \gamma c$ is the relativistic particle gyrofrequency corresponding to the unperturbed field $B_0$.
From that assumption it follows that the perpendicular diffusion coefficients (symmetric and anti-symmetric parts) are $\kappa_{\perp} = \kappa_B \Omega \tau_{\perp} / [1 + (\Omega \tau_{\perp})^2]$
and $\kappa_A = \kappa_B (\Omega \tau_{A})^2 / [1 + (\Omega \tau_{A})^2]$,
where $\tau_\perp$ and $\tau_A$ are the decorrelation time-scales ($\tau_\perp = \tau_A$ as assumed in the BAM model). 
Such a form of $\kappa_\perp$ generalizes 
the result of \citet{fjo74} down to low rigidities (below $4$ GV in interplanetary medium).
A comparison of the time-scale $\tau_\perp$ in the two forementioned works allows to relate 
the turbulent power at zero wavenumber to the intrinsic spatial coefficient diffusion 
for field line random walk. In the BAM model, it turns out also that 
$\kappa_{\perp} / \kappa_{\parallel} =  (\tau_{\perp}/\tau_{\parallel}  )/[1 + (\Omega \tau_{\perp})^2]$,
which collapses to the billiard ball scattering picture \citep{fg75} if $\tau = \tau_\parallel = \tau_\perp$.
In this paper we explore various regimes of the velocity autocorrelation.
The result is contrasted with the BAM model within broad ranges of $r_g/L_c$ and $\sigma^2$, 
where $r_g$ is the particle gyroradius, $ L_c$ the turbulence correlation length and $\sigma^2 = (\delta B/B_0)^2$ 
the magnetic turbulence normalized energy density.

Note that the most general form of the total perpendicular coefficient of diffusion (symmetric part) 
comprises both the field line meandering part, typically dominant contribution at low-rigidity ($\kappa_{MFL}$), 
and the departure from the local field line ($\kappa_{D}$), due to gradient/curvature drift: 
$\kappa_{\perp} = \kappa_{D} + \kappa_{MFL} v_{\parallel}$ \citep{fj11}.
We estimate in this paper the correlation time for the motion perpendicular to the direction of the average field as 
$\tau_{\perp} \simeq  (2/3)r_g ^2/ \kappa_{\perp}$, by using the numerically determined value of $\kappa_\perp$ 
(for the factor $(2/3)$ see Sect. \ref{fldec}).

\section{Numerical Method}\label{nummeth}

In a series of numerical experiments, we consider a population 
of charged test-particles gyrating in a magnetic field described as follows:
we assume a three-dimensional magnetic field of the form ${\bf B(x) = B}_0 + \delta {\bf B(x)}$, 
with an average component ${\bf B}_0 = B_0 {\bf e}_z$ and a random component 
${\bf \delta B} = {\bf \delta B} (x, y, z)$ having a zero mean ($\langle \delta {\bf B(x)} \rangle = 0$),
and has a turbulence correlation length $L_c$. We assume in the inertial range 
a scale-invariant, or Kolmogorov, power-spectrum 
in the three space-dimensions: $G(k) \propto k^{-\beta -2 }$, where $k$ is the wavenumber magnitude,
$\beta = 5/3$ is the one-dimensional power-law Kolmogorov index and the additional $2$ accounts 
for the dimensionality of the turbulence (for more details see appendix A).
The assumption of a static magnetic field is reasonable if the particle speed 
largely exceeds the Alfv\'en wave speed.

We perform numerical integration of equations of motion of charged particles 
combining the code used in \citet{fm08} with the prescription for the turbulence introduced in \citet{gj94}, 
and widely exploited in the last two decades in various astrophysics contexts: 
\citep{gj99, qmb02, bue08, fm08}. We determine the particle trajectory by numerically integrating the equation of motion 
using the Lorentz force determined at the instantaneous particle position (see Appendix A for details). 

We follow particle trajectories in a three-dimensional spatially unbounded region,
since particles escaped from a bounded computational domain 
would be removed from the simulation and could artificially modify
the estimate of the instantaneous diffusion coefficient. 
Charged particles are evolved in various realizations of the magnetic turbulence (see Appendix A for details).
The final time of our simulation runs is empirically determined
as the computational time where the asymptotic value for the diffusion coefficient is attained.

The relevant parameters are the ratio of the particle gyroradius 
to the turbulence correlations length, i.e., $r_g/L_c$, and the normalized turbulence energy density 
$\sigma ^2$. Therefore our treatment applies to energetic particles 
in various astrophysical environments: from the interplanetary space, 
to the supernova remnant shocks, and to the intergalactic medium.

\section{Diffusion coefficients}\label{diffcoeff}

\begin{figure}
\includegraphics[width=9cm]{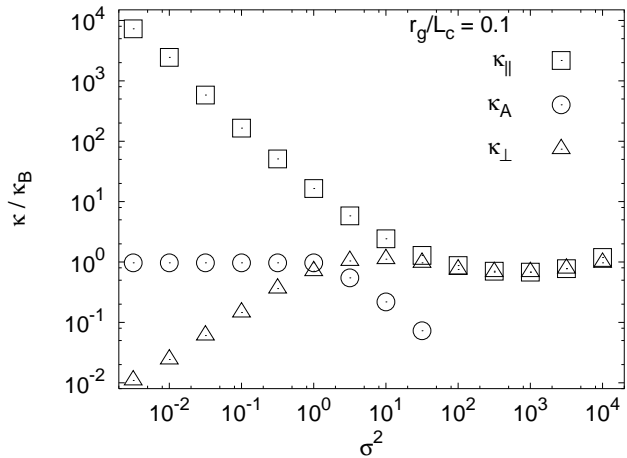}
\caption{Simulated parallel, anti-symmetric and perpendicular terms of the diffusion tensor, in units of $\kappa_B$, as a function of $\sigma^2$, for $r_g/L_c =0.1$. For weak turbulence, the classical transport theory prediction for the diffusion tensor apply (see also text). As expected, strong turbulence isotropize the particle transport ($\kappa_\perp \sim \kappa_\parallel$). A large power in the turbulence makes the field isotropic so no drift can be defined and $\kappa_A$ cannot be determined (see also \citet{gjk99})  \label{k_perp_3}}
\end{figure}

\begin{figure}
\includegraphics[width=9cm]{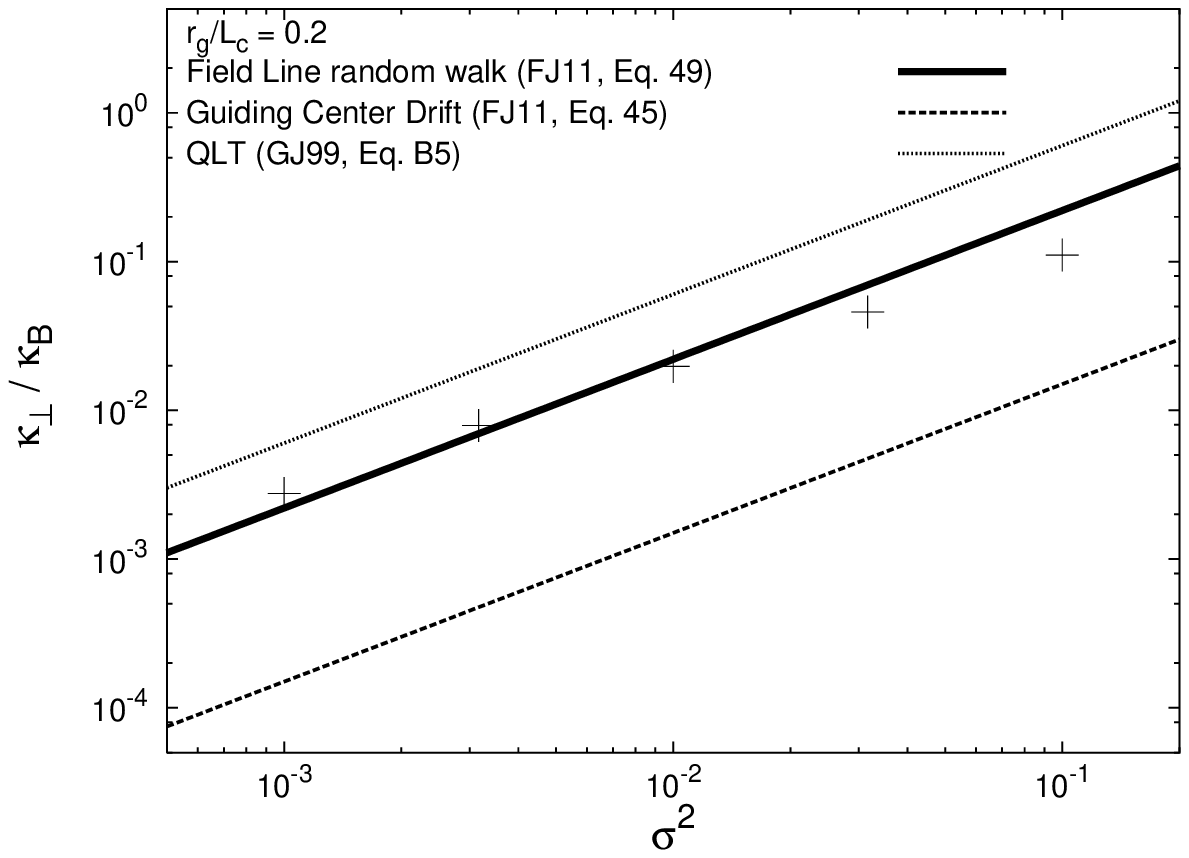}
\caption{Perpendicular diffusion coefficient, in units of $\kappa_B$, as a function of $\sigma^2$, for $r_g/L_c =0.2$. Simulated values are compared with the solid line and the dashed line, representing respectively the field line random walk and the departure from the local field line, or gradient/curvature drift diffusion coefficient, analytically found in  \citet{fj11}, here FJ11, Eqs. (49, 45), and the dotted line, representing the quasi-linear result, explicit in Eq. (B5) of \citet{gj99}, here GJ99.  \label{k_perp_drift}}
\end{figure}

\begin{figure}
\includegraphics[width=9cm]{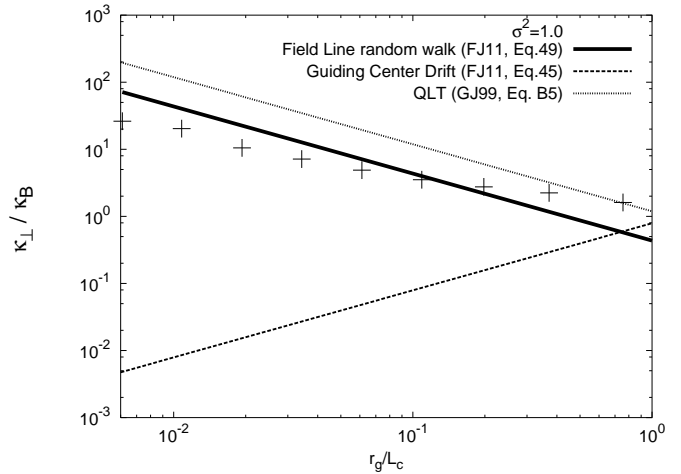}
\caption{Perpendicular diffusion coefficient, in units of $\kappa_B$, as a function of $r_g/L_c$, for $\sigma^2 = 1.0$. Simulated values are compared with the solid line and the dashed line, representing respectively the field line random walk and the departure from the local field line, or gradient/curvature drift diffusion coefficient, analytically found in  \citet{fj11}, here FJ11, Eqs. (49, 45), and the dotted line, representing the quasi-linear result, explicit in Eq. (B5) of \citet{gj99}, here GJ99. \label{k_perp_drift_2}}
\end{figure}

We report in this section the coefficients of diffusion parallel and perpendicular,
both symmetric and anti-symmetric, resulting from our numerical simulations (see Appendix B for details).
In Sect. \ref{velcorr} we will discuss the autocorrelation of the particle velocity 
making use of the diffusion coefficients computed in the present section 
to estimate the characteristic time-scale for the pitch-angle and the perpendicular scattering.

Fig.~(\ref{k_perp_3}) shows  the parallel ($\kappa_\parallel$), perpendicular ($\kappa_\perp$) and anti-symmetric term ($\kappa_A = \kappa_{xy} = - \kappa_{yx}$, see also Appendix B and Fig. (\ref{comp}), right panel, computed as $\kappa_{ij} = \langle v_i \Delta x_j \rangle$) of the diffusion tensor as a function of $\sigma^2$ at fixed particle energy conveniently scaled in units of Bohm diffusion coefficient $\kappa_B = v r_g /3$. In the weak turbulence limit, the behaviour of the diffusion tensor as a function of $\sigma^2$ ($\kappa_\perp \sim \sigma^2$ and $\kappa_\parallel \sim \sigma^{-2}$) is predicted by the standard quasi-linear theory for non-relativistic particle energies (see for example the derivation in the appendix of \citet{gj99}). For relativistic particles, such a power law dependence is unchanged.  The $\kappa_A$, known to be related to the gradient/curvature drift velocity, is constant as predicted by the classical scattering theory \citep{fg75}: in classical scattering theory $\kappa_A/\kappa_\parallel = (\lambda_\parallel / r_g) /(1+ (\lambda_\parallel / r_g)^2)$ at weak turbulence.
Thus, $\kappa_A / \kappa_B = (\kappa_\parallel / \kappa_B) / (1+(\kappa_\parallel / \kappa_B)^2) \rightarrow 1$ for $\sigma^2 \ll 1$, regardless of the particle energy (see also \citet{gjk99}). 

In the strong turbulence limit shown in Fig.~(\ref{k_perp_3}) the particle transport is isotropized ($\kappa_\perp = \kappa_\parallel$). This finding may be due to the particular choice of the turbulence power spectrum: different turbulence power spectrum can result in a different strong turbulence behaviour. The $\kappa_\perp$ and  $\kappa_\parallel$ merge to the same value at $\sigma_{\star} ^2 \sim 30$; the particle energy is not expected to affect much the value of $\sigma_{\star} ^2$. We notice that the convergence of $\kappa_\perp$ and  $\kappa_\parallel$ to $\kappa_B$ holds only for sufficiently large particle energy: for $r_g/L_c \ll 0.1$, the mean free path is much greater than the particle gyroradius so the Bohm diffusion underestimates the diffusion coefficients at strong turbulence.

Fig.~(\ref{k_perp_drift}) focuses on the $\sigma^2$-dependence of the $\kappa_\perp$ for $r_g/L_c = 0.2$ and $\sigma^2 \ll 1$. The result in \citet{fj11} is compared with the quasi-linear result, reported explicitly in Eq. (B5) of \citet{gj99}. The best agreement with the simulations is found using the approach in \citet{fj11}. For the 3D-isotropic turbulence, the only case considered in this paper, the gradient/curvature drift contribution is smaller by one order of magnitude than the field line random walk contribution.

Fig.~(\ref{k_perp_drift_2}) focuses on the dependence of $\kappa_\perp$ on $r_g/L_c$ for values of $r_g/L_c $ less than unity. The field-line random walk computed in \citet{fj11} provides the closest analytical result to the numerical simulations. Notice the increasingly relevant contribution of the gradient/curvature drift term. This shows that as the particle energy increases the departure from the local field line becomes relevant and the assumption that a particle follows the field line must fail, even for small $r_g/L_c$. Notice that  Fig.(\ref{k_perp_drift_2}) has been derived only for the case of 3D isotropic turbulence; it is not known how the gradient/curvature drift contribution to $\kappa_\perp$ will depend on other turbulence models.

\section{Velocity correlation}\label{velcorr}
In this section we study the time-dependence of the Lagrangian particle velocity autocorrelation $ \langle v_i(t) v_j(0)\rangle$. 
The resulting simulations are compared with models for pitch-angle scattering and perpendicular decorrelation. The decay time-scale of $\langle v_i(t) v_j(0)\rangle $ predicted in these models depends on the particle diffusion coefficients parallel and perpendicular to the average magnetic field, that we have estimated in the previous section.

\subsection{Pitch-angle scattering}

\begin{figure}
\includegraphics[width=9cm]{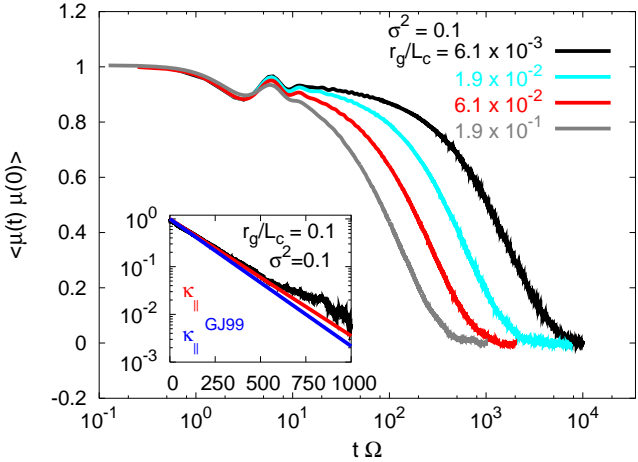}
\caption{Simulated pitch-angle correlation function for $\sigma^2=0.1$ and various particle energies as a function of $t\Omega$. 
Subpanel compares our simulations with a purely exponential decay having $\tau_\parallel$ as characteristic time-scale. At early-time, simulations (in black) agree with a purely exponential decay by making use of both values of $\kappa_\parallel = (v^2 /3) \tau_\parallel$, reported in this paper (in red) and in \citet{gj99} (in blue). \label{vvz-1}}
\end{figure}

\begin{figure}
\includegraphics[width=9cm]{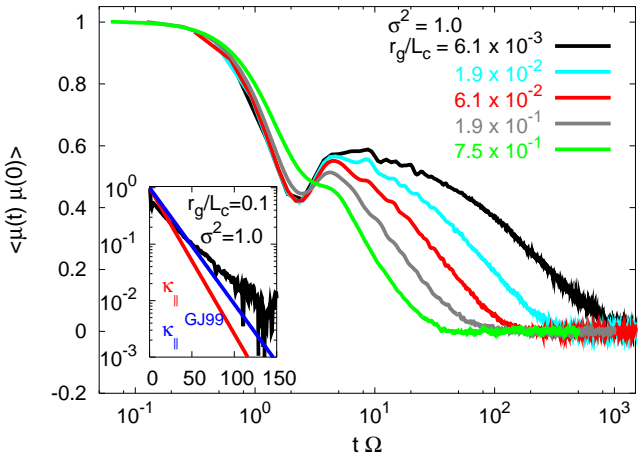}
\caption{Pitch-angle correlation function for $\sigma^2=1.0$ and various particle energies as a function of $t\Omega$. 
As in Fig. (\ref{vvz-1}), the subpanel focuses onto the early-time dependence. Intermediate energy density of turbulence can be quite satisfactorily reproduced by an exponential decay with $\tau_\parallel$ as characteristic time-scale (see also caption of Fig.(\ref{vvz-1})). \label{vvz0}}
\end{figure}

\begin{figure}
\includegraphics[width=9cm]{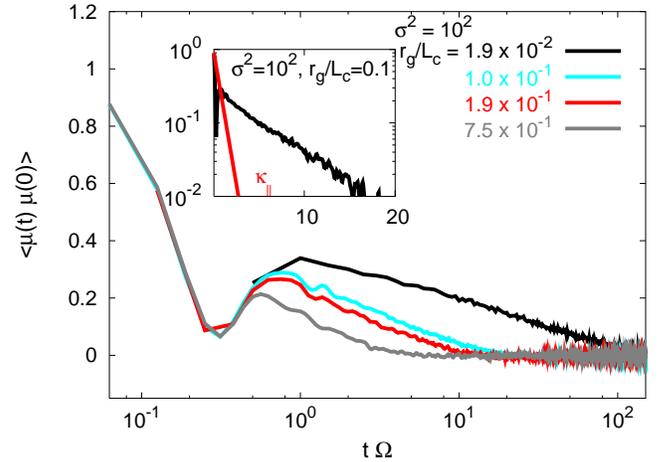}
\caption{Pitch-angle correlation function for $\sigma^2=10^2$ and various particle energies as a function of $t\Omega$. In the subpanel the simulation (in black) exhibits a temporal decay much slower than a simple exponential (in red). \label{vvz2}}
\end{figure}

In Figs. (\ref{vvz-1},\ref{vvz0},\ref{vvz2}), our simulations show that the pitch-angle correlation 
drops to zero as a function of time, or $t \Omega$, where $\Omega = e B_0/m \gamma c$ 
is the relativistic particle gyrofrequency corresponding to the unperturbed field $B_0$, 
for different values of $\sigma^2$ and for various particle energies: 
the stronger the turbulence and larger the particle energy, the faster the correlation decay even within a gyroperiod scale. 
The subpanels in Figs.(\ref{vvz-1},\ref{vvz0},\ref{vvz2}) compare the early-time decay from our simulations 
with the isotropic scattering form $e^{-t/\tau_\parallel}$ \citep{e74}.
The scattering time $\tau_{\parallel}$ is estimated by using $\tau_{\parallel} = (3/v^2) \kappa_{\parallel}$, 
where $\kappa_{\parallel} $ is computed in our simulation runs and compared with the values from \citet{gj99}. 
The subpanel inside Fig.(\ref{vvz-1}) confirms the exponential decay of the pitch-angle correlation $\sigma^2 \ll 1$. 
The bumps in the pitch-angle correlation at multiple integers of gyroperiod 
$t_g = 2\pi/\Omega$ suggest that, in its helicoidal trajectory, the particle velocity keeps 
higher order harmonics of its perturbed periodic motion (see also \citet{clp02}). 

Fig.(\ref{vvz0}) shows that for $r_g/L_c = 0 .75$ the bump at $t\Omega = 2\pi$ 
is smeared out because particles at such an energy ($1$ GeV in the interplanetary medium)
can travel a distance as large as $L_c$ experiencing therefore several scatterings within one gyroperiod
(compare also with the uppermost panel in Fig.(\ref{vvx_E})). 

The subpanel in Fig.(\ref{vvz2}) clearly shows for the first time directly in terms of $\langle \mu(0) \mu(t) \rangle$ 
that, for strong turbulence, the simulated correlation drops to zero significantly slower 
than $e^{-t/\tau_\parallel}$. Such a non-Markovian behaviour is found to hold independently on the particle rigidity. 
We have shown in Fig.(\ref{k_perp_3}) that $\kappa_\parallel$ (and therefore $\tau_\parallel$) saturates 
to a constant value for $\sigma^2 \gg 1$ at fixed energy; therefore, the deviation from
the Markovian scattering shown in the subpanel in Fig. (\ref{vvz2}) cannot be simply accounted for with an {\it ad hoc}
modification of the $\sigma^2$-dependence of $\tau_\parallel$. 
Within the quasi-linear limit, i.e., for small magnetic field fluctuations, \citet{f77} showed that the
weighted sum over all the characteristic times that expresses $\kappa_\parallel$ in Eq.(18) of \citet{e74} 
is equivalent to the time-integral up to infinity of the parallel velocity autocorrelation. 
The lowest order term of the Earl's series for $\kappa_\parallel$, corresponding to the case of purely isotropic scattering, 
can be obtained as time-integral of a simply exponential decay correlation function, 
with characteristic time given by $\tau_\parallel$. Higher order terms, that correspond to anisotropies
in the scattering, would produce a long-duration tail in the correlation similar to what we find in the
subpanel in Fig. (\ref{vvz2}) (see also Eq.(32) et seq. in \citet{f77}). 
We infer from Fig.(\ref{vvz2}) that enhanced turbulence produces a ``memory effect'' in the pitch-angle scattering 
comparable to the effect of the anisotropy terms in the weak turbulence. 
However, we do not perform a fit of our simulations due to the theoretical uncertainties underlying 
the higher-order characteristic time-scales in the Earl series.
We notice that a pure exponential $e^{-t/\tau_\parallel}$ underestimates the pitch angle correlation 
at early-time and therefore the asymptotic value of the diffusion coefficient $\kappa_\parallel$. 
Since this paper focuses on test particle simulations we will not develop an analytic discussion here.

\subsection{Field line decorrelation}\label{fldec}
 
\begin{figure}
\includegraphics[width=8.8cm]{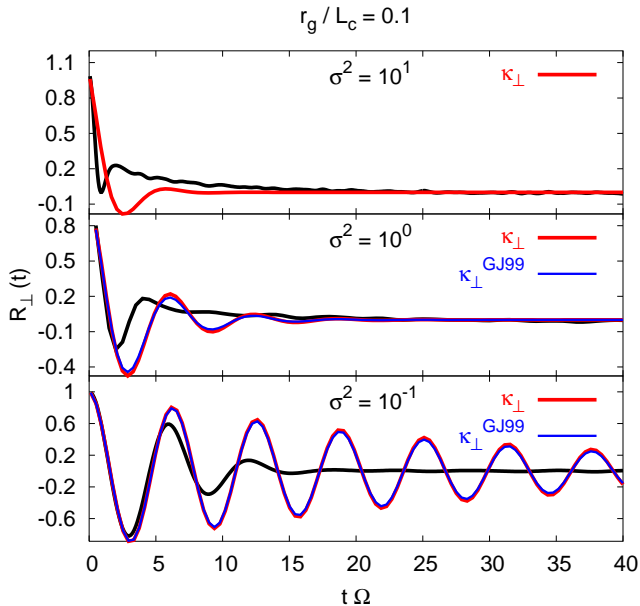}
\caption{Perpendicular velocity correlation function for $r_g/L_c = 0.1$ 
as a function of $t\Omega$. The three panels correspond 
to three values of turbulence energy density $\sigma^2$: $0.1, 1, 10$.
In every panel our simulations (in black) are compared with BAM model using 
the diffusion coefficients estimated both in the present paper (in red) and in \citet{gj99} (in blue). \label{vvx_varB}}
\end{figure}

\begin{figure}
\includegraphics[width=9cm]{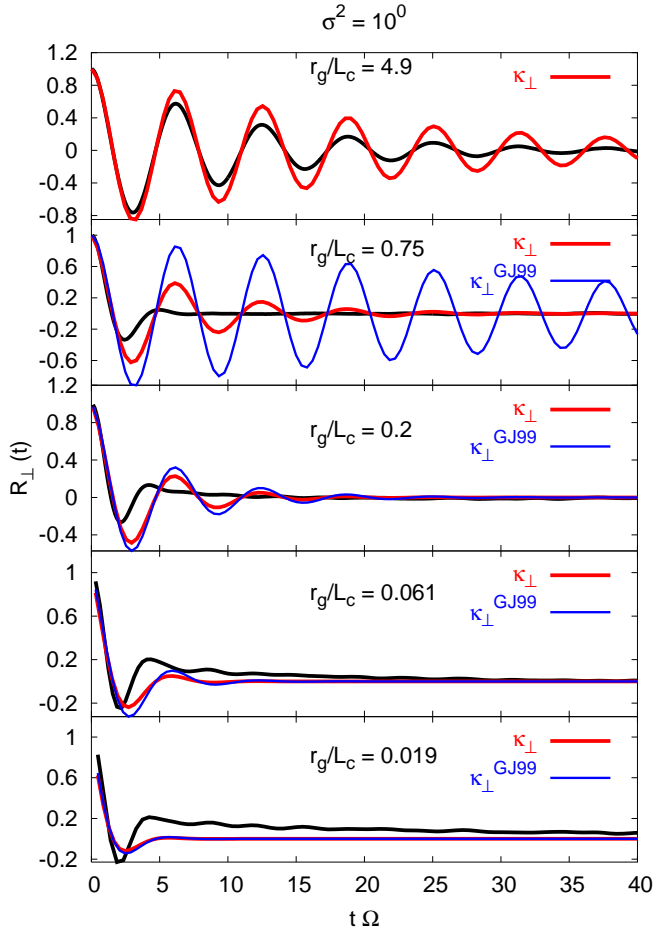}
\caption{Perpendicular velocity correlation function for $\sigma ^2=1.0$ as a function of $t\Omega$ corresponding
to five different values of particle energy ($r_g / L_c = 0.019, 0.061, 0.2, 0.75, 4.9$).
In every panel our simulations (in black) are compared with BAM model using 
the diffusion coefficients estimated both in the present paper (in red) and in \citet{gj99}  (in blue); 
see also caption of Fig.(\ref{vvx_varB}). \label{vvx_E}}
\end{figure}

\begin{figure}
\includegraphics[width=9cm]{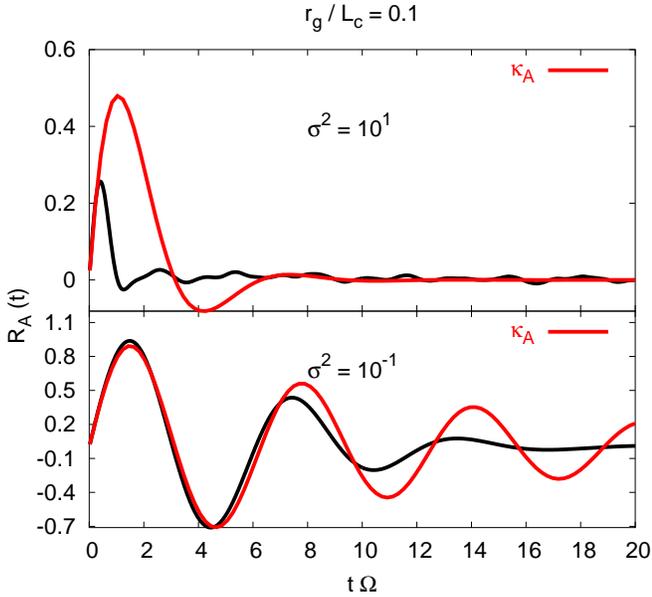}
\caption{Perpendicular antisymmetric velocity correlation function for $r_g /L_c = 0.1$ 
as a function of $t\Omega$ corresponding $\sigma^2 = 0.1, 10$.
In every panel our simulation (in black) is compared with BAM model (in red) using 
the diffusion coefficients estimated in the present paper (as illustrated in Appendix B).
\label{vvx_anti}}
\end{figure}

\begin{figure}
\includegraphics[width=9cm]{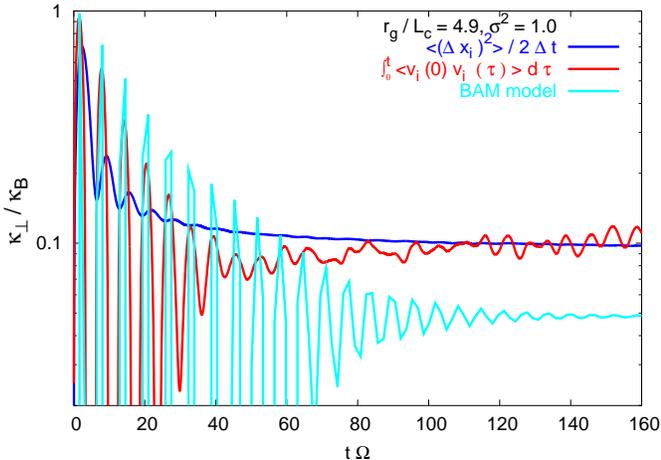}
\caption{Instantaneous perpendicular diffusion coefficient, in $\kappa_B$ units, comparing the
standard average square displacement (in blue) and the TGK (in red) methods with
the time integral of the $R_\perp (t)$ provided by the BAM model (in cyan). 
We chose here $r_g /L_c = 4.9$ and $\sigma^2 = 1.0$, in the BAM model regime validity. 
\label{vvx_comp_bm}}
\end{figure}

In this section we present our results for the perpendicular velocity autocorrelation on the gyroperiod scale.
Fig.(\ref{vvx_varB}) shows that the simulated autocorrelation deviates from the BAM model ansatz 
even for relatively small values of magnetic fluctuations and of $r_g/L_c$, 
within the expected regime of validity of the BAM model\footnote{An additional numerical factor is present in \citet{bm97}
to agree with the QLT limit: $\Omega \tau_\perp = (2/3) r_g/\kappa_{MFL}$ (analogous to Eq. 14 in their paper). However, the rate of decrease of the correlation does not depend significantly on the particular numerical factor used.
We use $\tau_\perp = (2/3) r_g^2 /\kappa_\perp$ to compare our simulations with the BAM model.}.
The two curves corresponding to $\kappa_\perp$ computed in this paper 
and to the value in \citet{gj99} agree in both the middle and lower panels.
For weak turbulence (lower panel), BAM model predicts 
a much less effective attenuation: the simulated correlation is completely smeared out 
beyond $t \Omega \sim 6\pi$, or $t \sim 3 t_g$. At intermediate fluctuation level (mid-panel),
the exponential suppression occurs faster; this is due to the fact that large fluctuations
increase the statistical value of the particle gyrofrequency ($\Omega = eB/m\gamma c$), 
computed as an ensemble-average, thus reducing the gyroradius ($r_g = v/\Omega$), 
i.e., the instantaneous radius of curvature of the particle trajectory; thus, the minimum in the velocity component
perpendicular to the average field is reached at an earlier time (see Fig.(\ref{vvx_varB}) mid-panel).
We note also that the increase of the gyrofrequency $\Omega$ at larger magnetic fluctuations
shortens the period of the oscillations, in absolute time.
These effects are enhanced at larger turbulence (upper panel in Fig.(\ref{vvx_varB})), 
where the oscillations are smeared out and the shape approaches a purely exponential decay
with a surprisingly fast drop.

The energy dependence of the $R_\perp (t)$ is depicted in Fig.(\ref{vvx_E}).
We find a good agreement between our simulations and the BAM model only at large particle energies 
($E \simeq 10$ GeV or $r_g / L_c = 4.9$ for interplanetary medium, uppermost panel), where the circular motion of the particle is weakly 
affected by the field line meandering, occurring at smaller scale.
However, as the rigidity is reduced, the particle motion resonates with a larger range of turbulence scales;
the velocity autocorrelation cannot be retained.

For completeness, Fig.(\ref{vvx_anti}) compares our simulations for $R_A (t)$ with the BAM model. 
Here $\tau_A$ is estimated as $\tau_A = \tau_\perp = (2/3) r_g^2/\kappa_\perp$.
Our simulations confirm the BAM ansatz for the velocity autocorrelation with the assumption $\tau_\perp = \tau_A$ 
only for weak turbulence within the gyroperiod scale ($t\sim 2 t_g$ for $\sigma^2 = 0.1$, 
compare Fig.(\ref{vvx_varB}), lower and upper panels).

Note that the instantaneous diffusion coefficient in the direction perpendicular 
to the average field is the time-integral of $\langle v_\perp (t) v_\perp (0)\rangle$, 
from zero to infinity. If the decay is sufficiently fast, as shown in Figs. (\ref{vvx_varB},\ref{vvx_E})
even for low values of $r_g/L_c$ and small fluctuations ($\sigma^2 \ll 1$), 
a substantial correction to the BAM modulated exponential is required. 
Currently no analytic models can describe such a rapid drop in the correlation.

Fig.(\ref{vvx_comp_bm}) compares the instantaneous $\kappa_\perp (t)$ 
from the BAM model ($\kappa_\perp (t) = \int_0 ^t d\xi R_\perp (\xi)$) with our simulations.
Clearly the BAM model is only applicable for long times due to sinusoidal factors, 
which make $R(t)$ assuming positive and negative values (see Fig.(\ref{vvx_comp_bm})).
We point out in this paper that the early-time oscillations in $R(t)$ cause the $\kappa_\perp$
underestimate previously found \citep{gj99}. Therefore, we conclude that a model 
for the perpendicular diffusion needs to ascertain an adequate description of the early-time propagation. 
On the other hand the drawbacks of a diffusion model can be identified 
through a study of the early-time propagation similarly to the study presented here.

\section{Some astrophysical applications}\label{astro}

In the preceding sections the charged particle motion has been described 
prior to the diffusion regime in terms of the particle velocity autocorrelation. 
This treatment applies to various astrophysical environments:
energetic particles ($E = 0.1 - 10^4$ MeV) in interplanetary space, 
at distance of $1$ AU from the Sun, gyrating in an ordered magnetic field $B_0 \sim 5$ nT,
with a turbulence $\sigma^2 \leq 1$ and a turbulence correlation length $L_c = 10^{-2}$ AU;
particles at energy $E = 5.1 \times 10^{13}$ eV to $E = 4.1 \times 10^{16}$ eV 
diffusing upstream or downstream of the non-relativistic shock
of supernova remnants, likely in an ordered field $B_0 = 3 \mu$G with an amplified turbulence $\sigma^2 \leq 10^4$,
and a turbulence correlation length $L_c \sim 3$ pc; 
particles at energy $E  =1.7\times 10^{14}$ eV to $E = 1.4\times 10^{17}$ eV transported in the turbulent galactic medium
with $B_0 = 3 \mu$G, $\sigma^2 \leq 1$ and $L_c \sim 10$ pc.

As an example of application of our simulations, we consider the problem of propagation
of energetic particles in the interplanetary medium detected by a spacecraft 
measuring the {\it in situ} magnetic-field. Consider a solar energetic particle, i.e., proton, 
with a kinetic energy $T = 15$ MeV in an approximately static solar wind Alfv\'enic perturbation,
released by the CME shock propagating from a gradual event.
At the location of particle release from the acceleration region,
if the turbulence can be described by a three-dimensional isotropic power spectrum,  
with $L_c = 10^{-3}$ AU and $B_0 = 5 $ nT ($r_g/L_c = 0.75$), 
Fig. \ref{vvx_E} shows that, due to perpendicular diffusion, the correlation drops on the gyroperiod scale 
($t_g = 2\pi/\Omega \sim 14$ sec). Likewise, if a GLE proton with $T = 10$ GeV is released 
in a turbulent interplanetary space with $L_c = 10^{-2}$ AU and $B_0 = 5 $ nT
($r_g/L_c = 4.9$), the decorrelation will drop on the scale of minutes ($t_g \sim 152$ sec).
Therefore, particle decorrelation from turbulent magnetic field lines needs to be considered 
in tracing the energetic particle trajectories in the interplanetary medium.
The anisotropic phase of the observed flux from solar particle events can be strongly affected.

The above result has also application to the galactic cosmic-rays below the ``knee'' 
of the cosmic-ray spectrum, thought to be accelerated 
at the supernova remnant shock, and escaping into the turbulent interstellar medium. 
At those shocks an efficient magnetic field amplification has been inferred 
up to values largely exceeding the Rankine-Hugoniot jump across the shock 
through various independent methods, e.g., from the shape of radio synchrotron spectra of energetic electrons \citep{re92} 
to $X-$ray rims in the remnant interior \citep{bv03}.
Various explanations have been proposed in the literature for such a large amplified magnetic field: 
instability by non-resonant cosmic-rays streaming upstream of the shock \citep{b04} 
or vortical turbulent motion seeding downstream magnetic field amplification \citep{gj07}.
In our simulations, assuming a large magnetic fluctuations, i.e., $(\delta B / B_0)^2 \sim 10^4$, 
$L_c = 3$ pc and  $B_0 = 1 \mu $G, an energetic proton with energy $E = 2.1 \times 10^{15}$ eV 
($r_g/L_c = 0.75$), will decorrelate on timescale of $2 t_g = 92 $ yrs 
(see for example Fig. \ref{vvx_E}, corresponding to $\sigma^2 =1.0$, panel with $r_g/L_c = 0.75$).
Our simulations do not include synchrotron energy losses, negligible for protons,
but likely to be relevant within a decorrelation time in the case of energetic electrons. 

\section{Summary and conclusion}\label{summconcl}

In this paper we have performed Monte-Carlo simulations of test particles 
gyrating in a static spatially
turbulent magnetic field with a three-dimensional isotropic power spectrum.
First, we have computed the dependence of the diffusion coefficients 
in the directions parallel and perpendicular to the average magnetic field
on the turbulence energy density. For weak turbulence, 
the prediction of our previous model for perpendicular transport based on the separation 
between the field line meandering and the gradient/curvature drift from the local field line
is found to be in better agreement with our numerical results than the standard quasi-linear theory.
For the particular power spectrum that we have considered and at fixed particle energy, 
the drift contribution is about one order of magnitude smaller than the contribution from field line meandering.
For strong turbulence, the diffusion tensor becomes isotropic.
We have also computed the dependence of the diffusion tensor on the particle energy,
specifically the ratio $r_g/L_c$. Also in this case the field-line meandering 
is found to be in better agreement with our numerical results than the standard quasi-linear theory.
The gradient/curvature drift term turns out to be important with the increase
of particle energy; a drift-dominated diffusion cannot be ruled out in different turbulent power spectra. 

Second, we computed the dependence of the particle velocity autocorrelation 
in the directions parallel and perpendicular to the average magnetic field
on the particle energy and on the turbulence energy density. 
The pitch-angle correlation drops exponentially in time for weak fluctuations, 
as predicted for the isotropic scattering case. We also found that for the case of strong turbulence 
the decay is not exponential. The deviation from the exponential decay in strong turbulence 
cannot be accounted for by current models. The perpendicular velocity autocorrelation decays faster than 
an exponentially modulated oscillation predicted by previous models. Even in the weak turbulence case 
and for gyroradius smaller than the turbulence correlation scale,
no significant correlation is found beyond three gyroperiods,
whereas the modulated exponential decays much slower.
Although in a strong turbulence the particle gyroradius cannot be uniquely defined,
our simulations show that the statistical effect of strong turbulence is 
reducing the instantaneous radius of curvature of the particle orbit; 
the correlation is lost within a fraction of gyroperiod.
However, for gyroradii larger than the correlation length the exponential modulation 
agrees with the simulations: on this scale the effect of the turbulence is not relevant 
and the particle pursues a quasi-helicoidal motion in a uniform field. 
However, for smaller gyroradii, the estimate of the decorrelation time and 
the magnitude shows the need of a new model.

We do not provide here a phenomenological fit or analytical model for the velocity autocorrelation. 
Our work shows that the underestimate of the diffusion coefficient previously found in 
the classical scattering theory can be explained in terms of lack of consistent model 
for the scattering and decorrelation times, and sheds light for the future investigations.
A model for the diffusion coefficient based on the velocity autocorrelation 
needs to provide accurate description of the early-time transport,
because the contributions to the diffusion beyond the gyroperiod scale average out to zero. 

\acknowledgments

It is a pleasure to acknowledge the fruitful discussions with J. R. Jokipii and J. K\'ota.
This work was supported by NSF grant ATM0447354 and by NASA grants NNX07AH19G and NNX10AF24G. 
FF thanks S. Dalla for useful correspondence.
FF wishes to acknowledge the hospitality of Observatory of Paris-Meudon (France),
where part of this work was completed.

,
\appendix
\section{Numerical set-up}

The global magnetic field is written as a sum of a background field $\vect{B}_0$,
constant and statistically uniform, and a turbulent field varying in space and independent on time.
The equation of motion of a test-particle with charge $e$ and mass
$m$ moving with velocity $\vect{v} (t)$ in a magnetic field $\vect{B}(\vect{x})$ is the Lorentz equation
\begin{equation}
\frac{d\vect{u}(t)}{dt} =  \vect{u}(t)
\times\vect{\Omega}(\vect{x})\;,
\label{lorentz}
\end{equation}
where $\vect{\Omega}(\vect{x})  = e \vect{B}(\vect{x})/(mc\gamma)$ with 
$\gamma = 1/ \sqrt{1-(v/c)^2}$ the Lorentz factor, $c$ the speed of light in vacuum
and where $\vect{u} (t) = \gamma \vect{v} (t)/c$.
We calculate the trajectory of the particle in a magnetic field
as a solution of Eq.(\ref{lorentz})
where $t$ is the time in the rest frame of the propagation region.
The quantity $\vect{\Omega}(\vect{x})$ in Eq.(\ref{lorentz}) is given by
$\vect{\Omega}(\vect{x}) = \vect{\Omega}_0+\delta \vect{\Omega}(\vect{x})$
where $\vect{\Omega}_0 \equiv (e/mc\gamma)
\vect{B}_0$, in terms of the background magnetic field $\vect{B}_0$,
and $\delta \vect{\Omega}(\vect{x})$ is the turbulent magnetic field.
We verified that the particle energy is conserved with a relative accuracy of $10^{-5}$.
We ignore any large-scale background electric fields.

The fluctuating field comprises of equal intensity circularly polarized left-handed and right-handed Alfven waves.
The procedure of building the turbulence calls for the random
generation of a given number $N$ of transverse waves $\vect{k}_i$, $i=1,..,N$ at every point of
physical space where the particle is found, each with a random amplitude, phase
and orientation. We use the form of the power spectrum provided in \citet{gj99}, 
Eq.s (3) -- (7). The wavenumber $k_i$ of the $N$ wave modes 
ranges from $k_{min} = 2\pi/L_{max}$ to $k_{max}= 2\pi/L_{min}$, 
spanning over four decades $L_{max}/L_{min} = 10^4$. The coherence scale $L_c$, 
or bend-over scale, for the turbulence power spectrum, is chosen such that $L_c = 10^2 L_{min}$. 
The values of $k_i$-magnitude are logarithmically equispaced, $\Delta k/ k =$ constant. 
We sampled the power spectrum over a discrete number of modes $N_m$. 
We choose $N_m ^* = 188$ modes of magnitude $k_i$; 
the resulting coefficients of diffusion do not depend significantly on the sampling resolution in $k-$space for $N_m > N_m^*$.

We inject a large number of particles at the same point of space, i.e., the origin of the coordinate system,
with randomly oriented velocity, but fixed energy. 
We integrate the Eq. (\ref{lorentz}) by using a time-step adjustable 
Burlisch-Stoer method \citep{p86}, which in our numerical solution of $2^{nd}$ order ODE resulted more accurate 
than a Runge-Kutta $5^{th}$ order. This approach is well suited to investigate 
the particle transport in turbulence. A different approach of computing 
different realizations of the turbulent magnetic field in every cell of a structured Cartesian grid 
is much more time-consuming. The latter approach would also require adapting 
the lattice spacing in order to maintain the same space resolution in physical space.

As in \citet{gj99} the randomization is performed over the initial particle velocity orientation 
and over the turbulence realization, shuffled every 50 particles.
Including the fluctuating field statistics ensures a meaningful comparison 
with a theoretically computed turbulence power spectrum,
which is by definition an average over an ensemble of field realizations. 

\section{Computation of the diffusion coefficients}

\begin{figure}
\includegraphics[width=9cm]{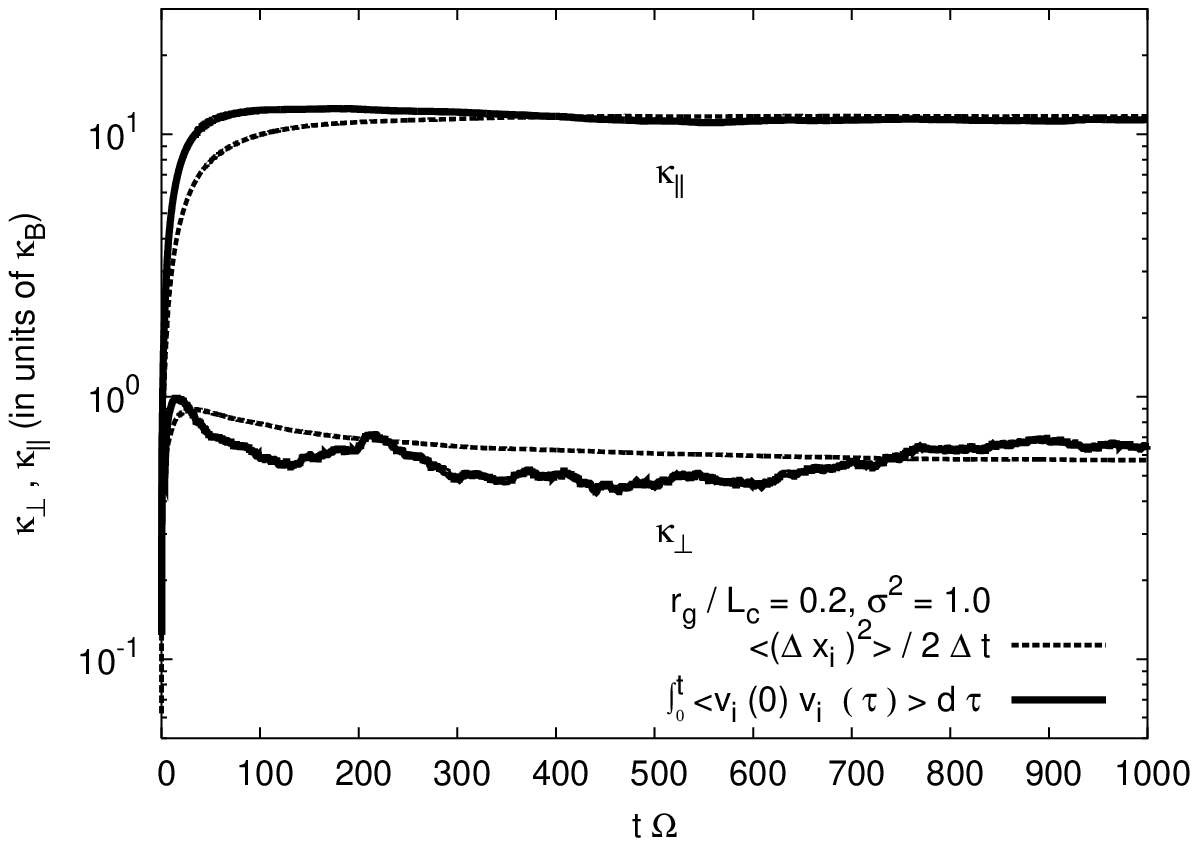}
\includegraphics[width=9cm]{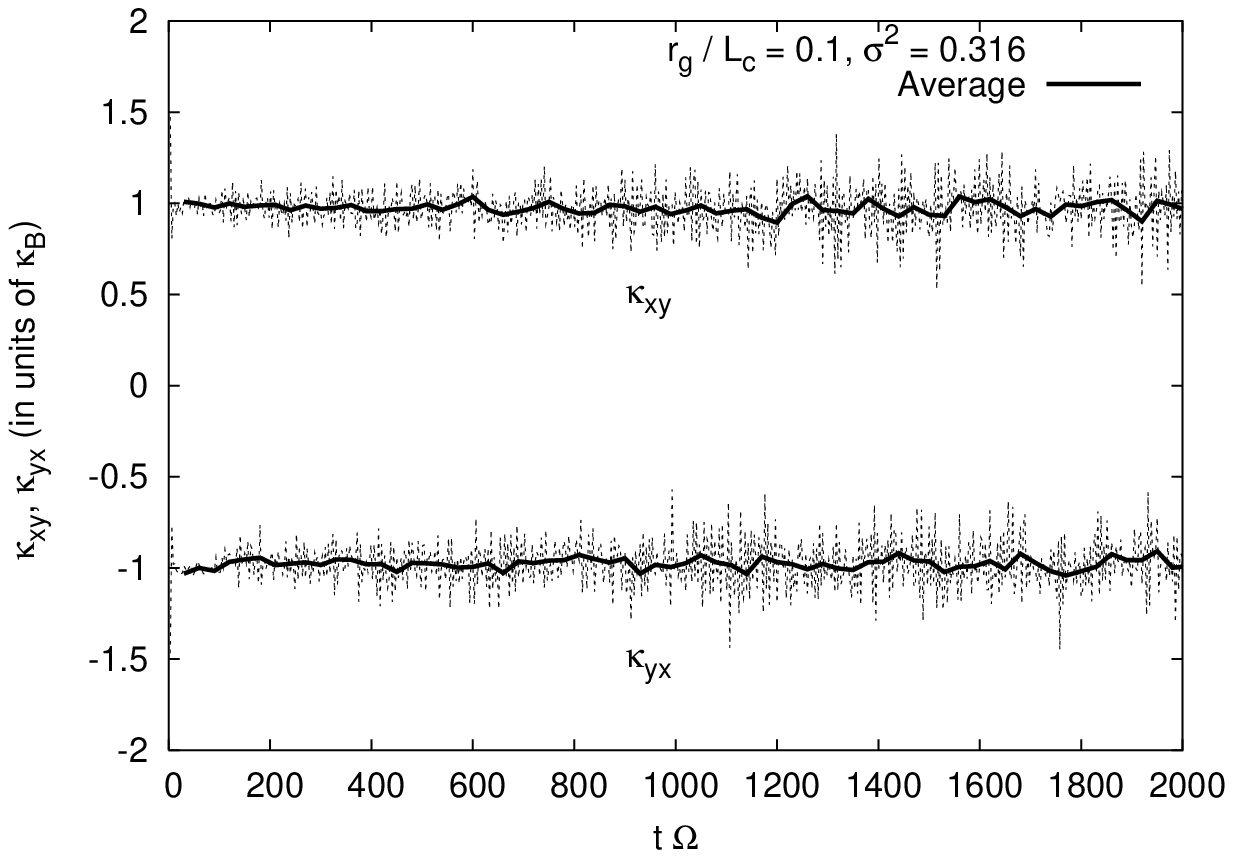}
\caption{{\bf Left} As a function of time, the diagram compares the coefficients of diffusion $\kappa_{\perp}$ and $\kappa_{\parallel}$ computed by using $\langle (\Delta x) ^2 \rangle  / 2t$ (in dashed) and  $\int_0 ^t  \langle  v_x (\xi) v_x (0) \rangle d\xi$, or TGK (in solid) for $r_g/L_c = 0.2$ and $\sigma^2 = 1.0$. If the TGK formula is used, the number of particles necessary to reduce numerical fluctuations is $N_p \simeq 50,000$. {\bf Right} As a function of time, the diagram compares the anti-diagonal terms of the diffusion tensor, i.e., $\kappa_{xy}$ and $\kappa_{yx}$, in units of $\kappa_B$, anti-symmetric as expected ($\kappa_{xy} = - \kappa_{yx}$). Here $r_g/L_c = 0.1$, $\sigma^2 = 0.316$ and the number of particles used is $N_p =50,000$. We draw the time-averages to clearly show the asymptotic value. \label{comp}}
\end{figure}

We computed the parallel ($\kappa_\parallel$) and symmetric perpendicular ($\kappa_\perp$) 
coefficients of diffusion in the previously specified turbulence using two different methods.
The first method uses the formula for the instantaneous coefficient diffusion
based on the ensemble-averaged mean square displacement:
\begin{equation}
d_{ij} ^{I} (t) = {1\over 2 N_p} \sum_{n=1} ^{N_p} \frac{(x_i(t) - x_i(t_0)) (x_j(t) - x_j(t_0))}{t-t_0}  
\label{coeff1}
\end{equation}
where $N_p$ is the total number of particles (the turbulence realization being shuffled every $50$ particles), 
$x_i (t)$ the particle trajectory along the i-th space coordinate and $t_0$ injection time. 

The second method involves the particle velocity autocorrelation given by the TGK formula:
\begin{equation}
d_{ij} ^{II} (t) =  \frac{1}{2} \frac{d}{d t} { \langle \Delta x_i \Delta x_j \rangle } = \int_0^{ t} d\xi  \langle v_i(\xi) v_j(0)\rangle \, ,
\label{coeff2}
\end{equation}
where $v_k (t)$ is particle velocity along the space coordinate $x_k$.
The last equivalence is valid for fluctuating velocity statistically homogeneous for sufficiently long time.

In both cases the coefficient of diffusion is estimated as the apparent asymptotic value 
in the diagram representing our numerical simulations (see Fig. (\ref{comp}), Left panel), 
that is attained after a large number of scatterings off the magnetic disturbance: 
$d_{ii} (t) \rightarrow \kappa_{ii}$ for $t \rightarrow \infty$.
For $r_g \ll L_c$ it takes many gyroperiods before the force acting upon the particle decorrelate from the initial force.
For $r_g \gtrsim L_c$, the force decorrelates much earlier, allowing eventually the particle
to gyrate around a field line different from the field line it was released on. 
Thus the diffusion time-scale ranges from $10^4 t \Omega$ (for $r_g \ll L_c$) to $10^2 t \Omega$ (for $r_g \gtrsim L_c$).

By using the first method (Eq.\ref{coeff1}), we found results numerically convergent and consistent with \citet{gj99} 
by injecting only $N_p = 2,500$ particles. However, the second method (Eq.\ref{coeff2}), due to the large numerical fluctuations around zero 
of the velocity correlation at $t \gg \Omega^{-1}$, 
requires a much larger number of particles, i.e., $N_p \simeq 50,000$,
for the results of the two methods to be comparable (see Fig.(\ref{comp}), Left panel).
The chosen number of particles compromises between reducing the computation time and smearing out 
the numerical fluctuations in the tail at $t \Omega > 10 -10^2$ in Figs. (\ref{vvz-1},\ref{vvz0},\ref{vvz2}). 

Equations (\ref{coeff1}, \ref{coeff2}) are symmetric in the indexes $i-j$, therefore not suitable to define the anti-symmetric part
of the diffusion tensor. We follow here the argument in \citet{gjk99}, valid for a nearly isotropic population 
of non-relativistic particles streaming without convection, to write the anti-symmetric part of the diffusion tensor: 
$\kappa_{ij} = \langle v_i \Delta x_j \rangle$, where $v_i$ is the velocity along the $i$ direction 
and $\Delta x_j$ the space displacement along the $j$ direction. In Fig.(\ref{comp}), Right panel, the averages of $\kappa_{xy}$
and $\kappa_{yx} = - \kappa_{xy}$ are plotted as a function of time, in order to smear out the large fluctuation 
due to velocity oscillations.

\clearpage

\end{document}